\documentclass{vgtc}                          

\graphicspath{{figures/}{pictures/}{images/}{./}} 

\usepackage{times}                     

\usepackage{tabu}                      
\usepackage{booktabs}                  
\usepackage{lipsum}                    
\usepackage{mwe}                       

\usepackage{mathptmx}                  

\onlineid{0}

\vgtccategory{Research}

\vgtcinsertpkg

\title{Towards Understanding the Impact of Guidance in Data Visualization Systems for Domain Experts}

\author{Sherry Qiu\thanks{e-mail: sherry.qiu@yale.edu} %
\and Holly Rushmeier\thanks{e-mail: holly.rushmeier@yale.edu} %
\and Kim R.M. Blenman\thanks{e-mail: kim.blenman@yale.edu}}
\affiliation{\scriptsize Yale University}

\abstract{
    Guided data visualization systems are highly useful for domain experts to highlight important trends in their large-scale and complex datasets. 
    However, more work is needed to understand the impact of guidance on interpreting data visualizations as well as on the resulting use of visualizations when communicating insights. We conducted two user studies with domain experts and found that experts benefit from a guided coarse-to-fine structure when using data visualization systems, as this is the same structure in which they communicate findings.
} 


\begin{document}


\firstsection{Introduction}

\maketitle
Guidance for data visualization systems is the support provided by the system to help resolve the user's "knowledge gaps" \cite{7534883}. Guidance helps domain experts detect trends and identify important areas of the data so they can generate insights and new knowledge \cite{9645173}. However, the workflow built into visualization systems is often unfamiliar to experts \cite{STOIBER202268}. 

Visualizations also play a role in the communication of findings. Data visualization systems exist in various domains \cite{filipov2023timelighting, leite2020hermes, preim2020survey}, however, little work has been done to understand how guidance affects the use of visualizations when communicating findings.

In this paper we verify the need for guidance in data visualization systems and storytelling of findings, choosing proteomics as our use case. 
The current proteomics analysis process is unstructured and involves various packages to generate visualizations. 
No research has been done to determine how the visualizations should be interpreted together. 
In this case the knowledge gap is unknown targets when coming to new conclusions about the dataset.
We conducted two studies with 11 domain experts consisting of end user biologists, computational biologists, and clinicians. Due to the expertise required to understand the data, it is common to recruit only a half a dozen or so experts for studies \cite{li2023spectrumva, 9645173}. We present findings that shed light on the impact of guidance on experts' interpretation of data visualizations and subsequent storytelling, and emphasize incorporating guidance into systems for developers.

\section{Study Design}

\subsection{Visualization Interface}

\begin{figure}[t]
    \centering
    \includegraphics[width=\linewidth]{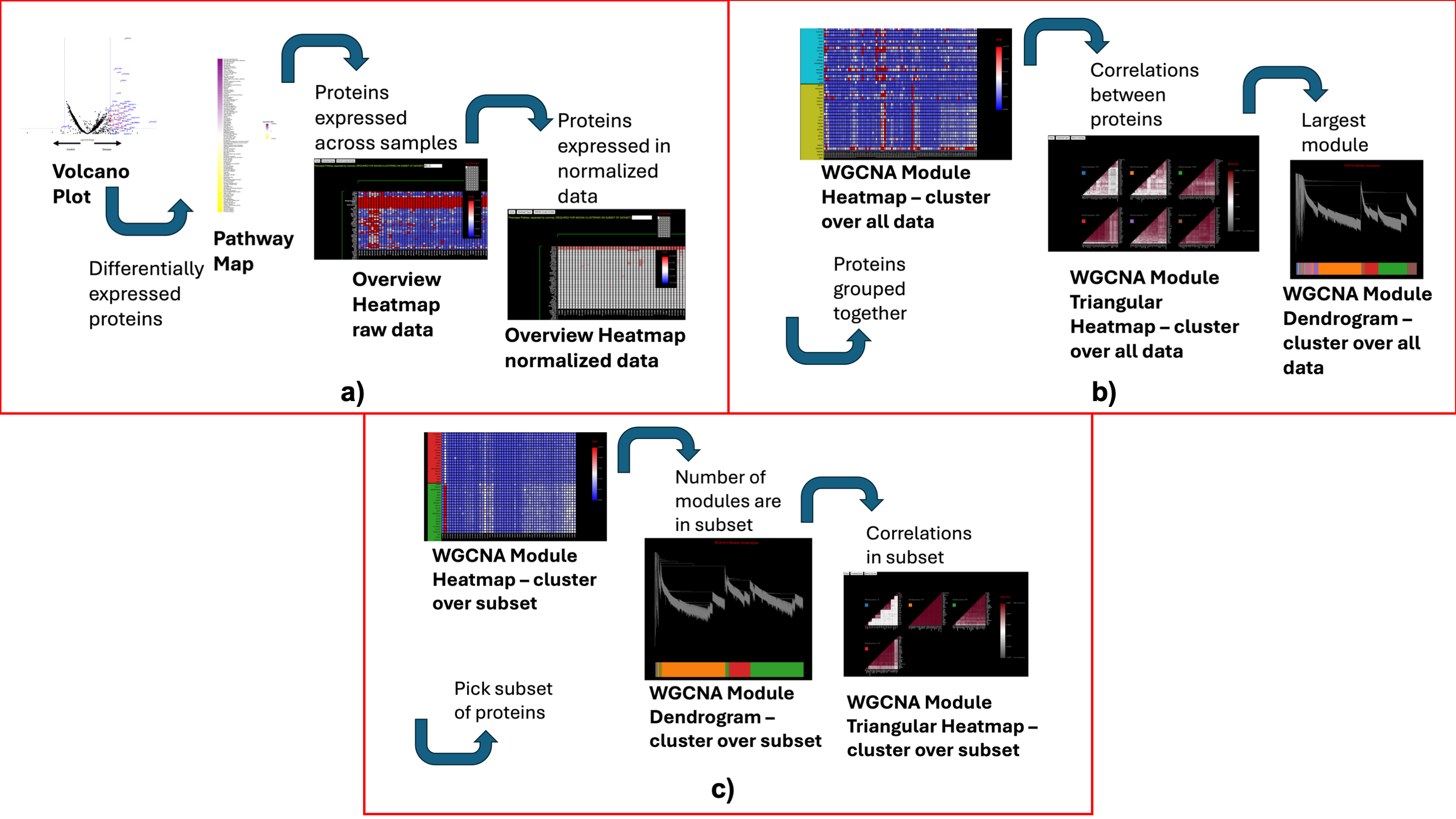}
    \caption{\label{fig:story} Panels a), b), and c) together demonstrate use of \textit{visAPPprot} visualizations to communicate findings about input dataset.
    }
\end{figure}

To conduct our studies, we developed a coarse-to-fine guided data visualization system \textit{visAPPprot}, based on Shneiderman's mantra \cite{545307}. 
The Volcano Plot, Pathway Map, and Overview Heatmap are common to proteomics and provide high-level impressions of the dataset. The WGCNA clustering technique \cite{langfelder2008wgcna} displays detailed information about modules (data subsets) and entity relationships. Fig. ~\ref{fig:story} shows our visualizations used in storytelling and our supplemental video explains the interactive features.

\subsection{Study Procedure}
We provided the same expression matrix of 89 patients (54 Disease; 35 Control) and 15927 proteins per patient for all participants. They recorded their screen with Zoom and completed a 5-point Likert scale survey on their impressions of the visualizations, with 1 = "Strongly disagree" and 5 = "Strongly agree".

\paragraph{Formative Study}

We divided participants into a guided and an unguided cohort and allowed them 90 minutes on a dedicated Windows 10 machine. The guided cohort received instructions with labeled diagrams directing them to complete 21 tasks and to view the visualizations in a coarse-to-fine order: 1) Volcano Plot, 2) Pathway Map, 3) Overview Heatmap, 4) WGCNA clustering on entire dataset, 5) WGCNA clustering on subset. The unguided cohort received documentation to view visualizations but no instructions on visualization order. We also asked participants to compare our dynamic heatmap with the standard proteomics static heatmaps.

\paragraph{Further Exploration}

Two tasks were assigned to both cohorts: storytelling and insights. 
For storytelling, each participant was asked to order the visualizations to tell a story about their findings and explain their ordering.
For insights, each participant was asked to list insights they gained into the dataset for each visualization.

\section{Findings}

\begin{figure}[t]
    \centering
    \includegraphics[width=8cm,height=8cm,keepaspectratio]{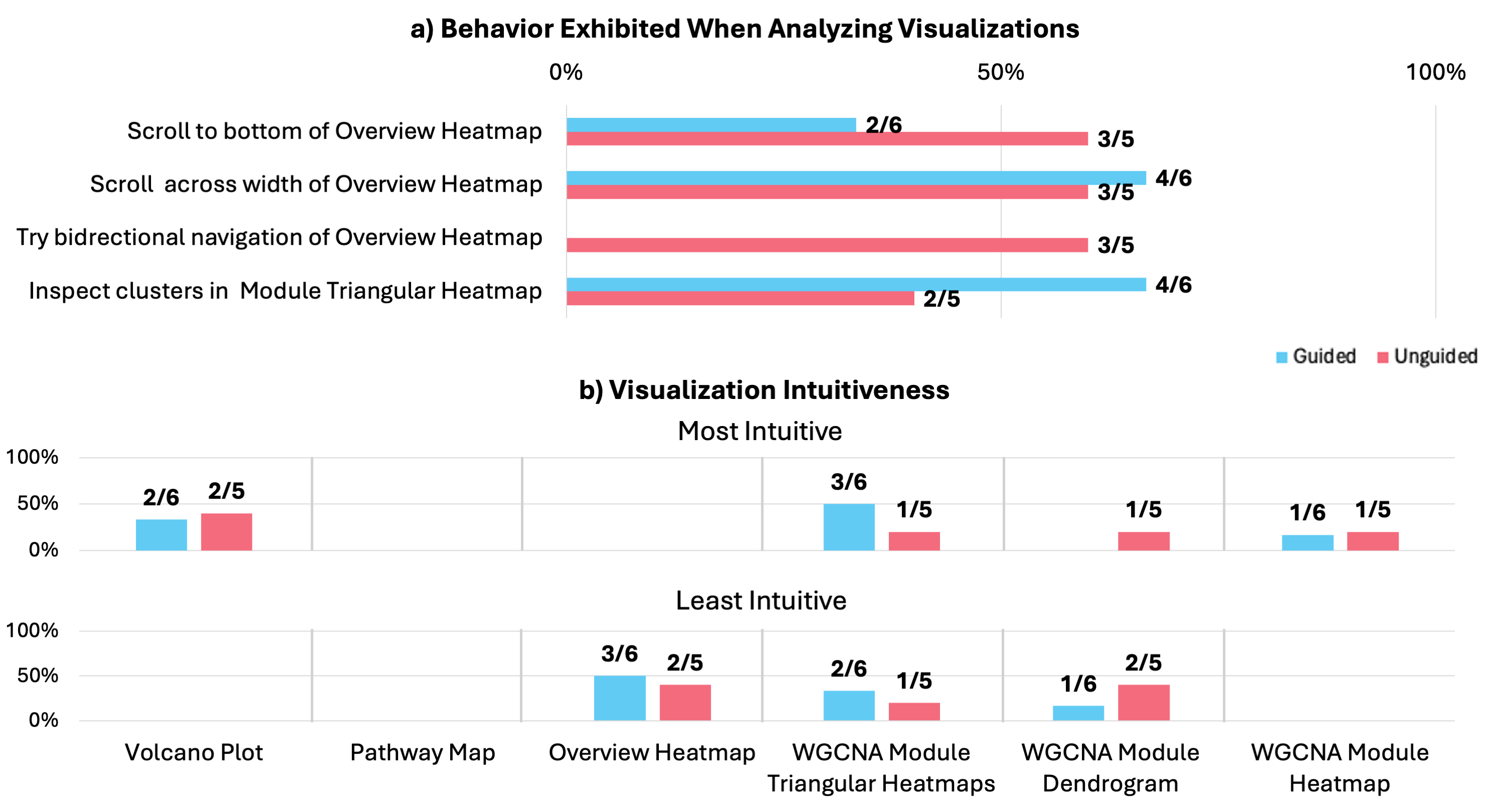}
    \caption{\label{fig:behavior_intuitive} Top: Behaviors exhibited by participants. Bottom: Most and least intuitive visualizations.}
\end{figure}

\begin{figure}[t]
    \centering
    \includegraphics[width=\linewidth]{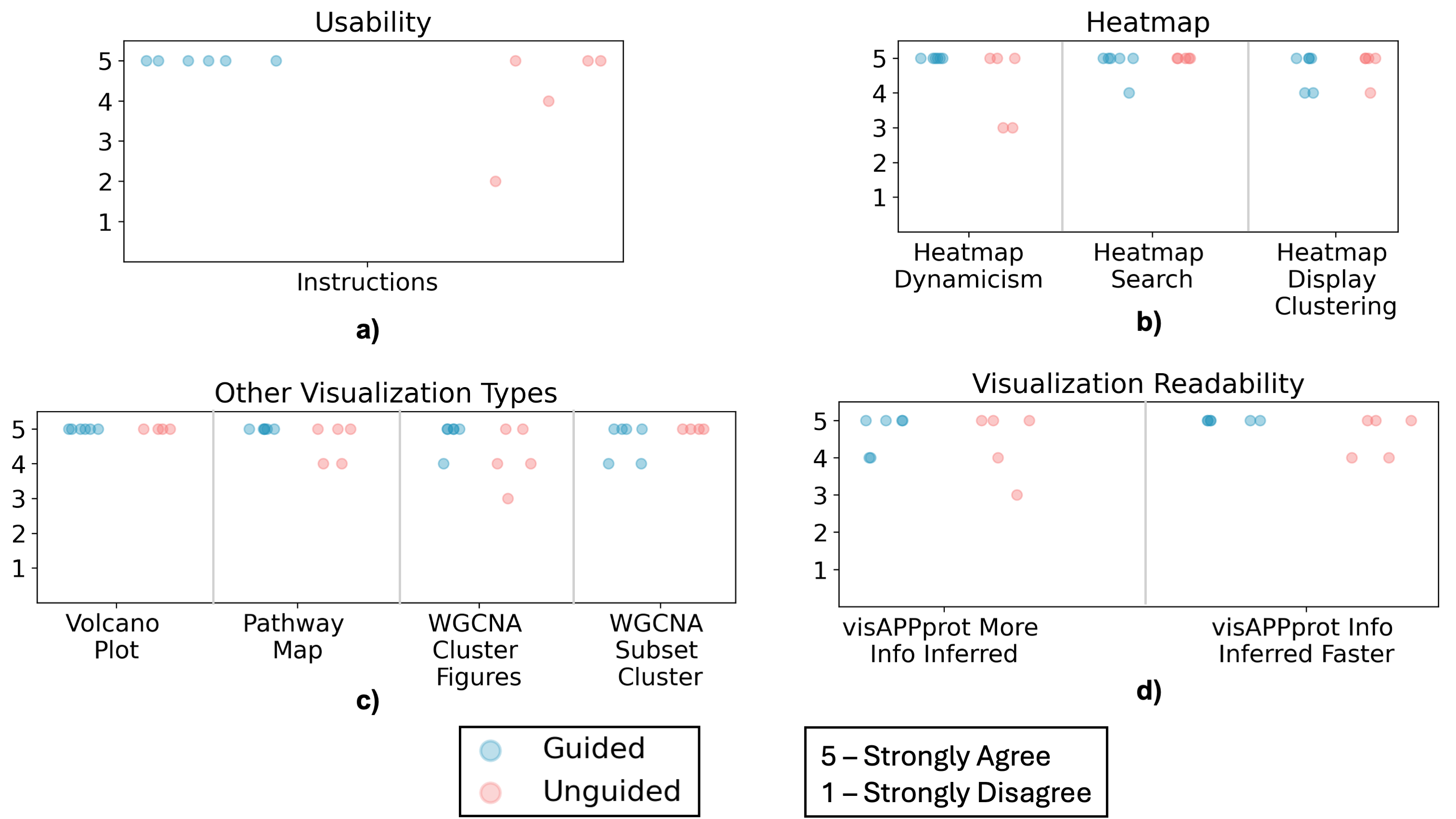}
    \caption{\label{fig:vis_understanding} Visualization interpretation by guided and unguided cohort, separated by category and plotted by individual participant results.
    }
\end{figure}

Overall, the unguided cohort had more difficulty understanding visualizations than the guided cohort and more trouble examining detailed information. The majority of participants developed coarse-to-fine storytelling structures. Applying open and axial coding to the insights revealed themes indicating areas of guidance participants valued: overview of the data, groups of entities, and individual entities. Results from our study are shown in Figs. ~\ref{fig:behavior_intuitive} and ~\ref{fig:vis_understanding}. We summarize key findings in the following paragraphs. 

\textbf{Guidance is needed to orient users in detecting high level trends.}
The unguided cohort was eager to see high-level trends in the dataset. Over half of the unguided participants tried to view the entire heatmap quickly through attempting bidirectional navigation of the marker in the navigation map to scroll the heatmap, while none of the guided participants exhibited this behavior.
The unguided participants explored the heatmap before viewing the other high-level visualizations, which would have provided more context for the heatmap. This result indicates system designers need to understand and provide guidance on the order of viewing visualizations as an indirect answer to the knowledge gap, intending to help users see high-level insights about the dataset \cite{7534883}. 

\textbf{Guidance is needed to orient users in exploring detailed information.} 
The unguided cohort took longer to view the subset figures and lacked direction in their exploration compared to the guided cohort. The unguided cohort was less likely to inspect clusters within each module 
and found the Module Triangular Heatmaps and Module Dendrogram less intuitive than the guided cohort. 
This suggests that system designers should integrate guidance as an indirect answer to the knowledge gap by encouraging experts to understand relationships within subsets of the dataset \cite{7534883}.

\textbf{Dynamic and interactive visualizations and visualizations organized by and supplemented with detailed information are more effective for displaying large datasets.} The Overview Heatmap maintains the legend, axes, and navigation map with location marker in view as the user scrolls. This encouraged participants to scroll across the width of the heatmap to examine specific protein expression.
In contrast, no participant viewed specific protein information on the static heatmap. Both cohorts still voted the Overview Heatmap one of the "least intuitive" visualizations.
Over half of all participants did not scroll vertically through the Overview Heatmap, 
suggesting that an overwhelming amount of information was displayed. In contrast, the WGCNA Module Heatmap was easily understood. 
The Module Heatmap displayed the same data as the Overview Heatmap but was organized by color-coded modules.
This suggests that system designers should incorporate guidance through dynamicism to introduce contextual information in the user's view and guidance through organization by detailed information to help experts better understand visualizations. 

\textbf{Participants communicated their findings in a coarse-to-fine structure.} When ordering stories, participants stated that they wanted a "big picture" and "broad view of all data" for the first visualization. For latter visualizations, participants wanted to show "relationships between proteins" and "groups of proteins". Almost all participants agreed that guided analysis would be helpful for storytelling, indicating that system designers should incorporate coarse-to-fine guidance in analysis to help experts communicate findings.

\section{Conclusion}

We present an evaluation of the impact of guidance in data visualization systems on domain experts' interpretation of visualizations and use of visualizations when communicating findings. 
We found that coarse-to-fine guidance for analyzing visualizations is effective for interpreting visualizations and subsequent storytelling of insights, where domain expertise is required distinguish between overview visualizations and visualizations showcasing detailed information.
We acknowledge that our limited sample size and qualitative results can introduce bias in our findings.

Our findings motivate the need to better understand how guidance affects visualizations systems as a whole, instead of individual visualizations. 
We reveal a need for a continuous dialogue between domain experts and developers, and in future research we will utilize our findings to create intuitive guided visualization systems.

\bibliographystyle{abbrv-doi}

\bibliography{template}

\begin{thebibliography}{1}

\bibitem{7534883}
D.~Ceneda, T.~Gschwandtner, T.~May, S.~Miksch, H.-J. Schulz, M.~Streit, and
  C.~Tominski.
\newblock Characterizing guidance in visual analytics.
\newblock {\em IEEE Transactions on Visualization and Computer Graphics},
  23(1):111--120, 2017. doi: {{%
10\hspace{.1pt}\discretionary{.}{%
}{.}\hspace{.4pt}1109\discretionary{/}{%
}{/}TVCG\hspace{.1pt}\discretionary{.}{%
}{.}\hspace{.4pt}2016\hspace{.1pt}\discretionary{.}{%
}{.}\hspace{.4pt}2598468}}


\bibitem{filipov2023timelighting}
V.~Filipov, D.~Ceneda, D.~Archambault, and A.~Arleo.
\newblock Timelighting: Guidance-enhanced exploration of 2d projections of
  temporal graphs.
\newblock In {\em International Symposium on Graph Drawing and Network
  Visualization}, pp. 231--245. Springer, 2023.

\bibitem{langfelder2008wgcna}
P.~Langfelder and S.~Horvath.
\newblock {WGCNA}: an r package for weighted correlation network analysis.
\newblock {\em BMC bioinformatics}, 9:1--13, 2008.

\bibitem{leite2020hermes}
R.~A. Leite, A.~Arleo, J.~Sorger, T.~Gschwandtner, and S.~Miksch.
\newblock Hermes: Guidance-enriched visual analytics for economic network
  exploration.
\newblock {\em Visual Informatics}, 4(4):11--22, 2020.

\bibitem{li2023spectrumva}
J.~Li, C.~Lai, Y.~Wang, A.~Luo, and X.~Yuan.
\newblock Spectrumva: Visual analysis of astronomical spectra for facilitating
  classification inspection.
\newblock {\em IEEE Transactions on Visualization and Computer Graphics}, 2023.

\bibitem{9645173}
M.~Meuschke, U.~Niemann, B.~Behrendt, M.~Gutberlet, B.~Preim, and K.~Lawonn.
\newblock Gucci - guided cardiac cohort investigation of blood flow data.
\newblock {\em IEEE Transactions on Visualization and Computer Graphics},
  29(03):1876--1892, mar 2023. doi: {{%
10\hspace{.1pt}\discretionary{.}{%
}{.}\hspace{.4pt}1109\discretionary{/}{%
}{/}TVCG\hspace{.1pt}\discretionary{.}{%
}{.}\hspace{.4pt}2021\hspace{.1pt}\discretionary{.}{%
}{.}\hspace{.4pt}3134083}}


\bibitem{preim2020survey}
B.~Preim and K.~Lawonn.
\newblock A survey of visual analytics for public health.
\newblock In {\em Computer Graphics Forum}, vol.~39, pp. 543--580. Wiley Online
  Library, 2020.

\bibitem{545307}
B.~Shneiderman.
\newblock The eyes have it: a task by data type taxonomy for information
  visualizations.
\newblock In {\em Proceedings 1996 IEEE Symposium on Visual Languages}, pp.
  336--343, 1996. doi: {{%
10\hspace{.1pt}\discretionary{.}{%
}{.}\hspace{.4pt}1109\discretionary{/}{%
}{/}VL\hspace{.1pt}\discretionary{.}{%
}{.}\hspace{.4pt}1996\hspace{.1pt}\discretionary{.}{%
}{.}\hspace{.4pt}545307}}


\bibitem{STOIBER202268}
C.~Stoiber, D.~Ceneda, M.~Wagner, V.~Schetinger, T.~Gschwandtner, M.~Streit,
  S.~Miksch, and W.~Aigner.
\newblock Perspectives of visualization onboarding and guidance in va.
\newblock {\em Visual Informatics}, 6(1):68--83, 2022. doi: {{%
10\hspace{.1pt}\discretionary{.}{%
}{.}\hspace{.4pt}1016\discretionary{/}{%
}{/}j\hspace{.1pt}\discretionary{.}{%
}{.}\hspace{.4pt}visinf\hspace{.1pt}\discretionary{.}{%
}{.}\hspace{.4pt}2022\hspace{.1pt}\discretionary{.}{%
}{.}\hspace{.4pt}02\hspace{.1pt}\discretionary{.}{%
}{.}\hspace{.4pt}005}}


\end{thebibliography}
\end{document}